\documentclass[12pt]{article}

\usepackage{amsmath,array,amssymb,cite} 

\def\elabel#1{\label{#1}}

\DeclareSymbolFont{AMSb}{U}{msb}{m}{n}
\DeclareSymbolFontAlphabet{\Bbb}{AMSb}

\newcommand{\aD}{{\dot\alpha}}

\newcommand{\bD}{{\dot\beta}}
\newcommand{\U}{{U}}
\newcommand{\SU}{{SU}}

\newcommand{\BL}{{\bf L}}

\def\com{X}

\def\sfc{\hat\Omega}

\def\N{{\cal N}}
\def\det{{\rm det}}

\def\M{{\cal M}}
\def\tr{{\rm tr}}
\def\trtwo{\tr^{}_2\,}
\def\Mbar{\bar{\cal M}}
\def\dalpha{{\dot\alpha}}
\def\dbeta{{\dot\beta}}

\def\Skinst{S^k_{\rm inst}}
\def\dmuphys{d\mu^{k}_{\rm phys}}
\def\Skquad{S^k_{\rm quad}}

\def\wbar{\bar w}
\def\mubar{\bar\mu}
\def\abar{\bar a}

\def\mubar{\bar\mu}

\def\Skinst{S^k_{\rm inst}}
\def\dmuphys{d\mu^k_{\rm phys}}
\def\N{{\cal N}}

\def\Skquad{S^k_{\rm quad}}

\def\Skinst{S^k_{\rm inst}}
\def\sst{\scriptscriptstyle}

\def\Mbar{\bar{\M}}

\def\K{{\cal K}}

\def\D{{\cal D}}
\def\Dbarslash{\,\,{\raise.15ex\hbox{/}\mkern-12mu {\bar\D}}}
\def\delslash{\,\,{\raise.15ex\hbox{/}\mkern-9mu \partial}}
\def\Dslash{\,\,{\raise.15ex\hbox{/}\mkern-12mu \D}}

\def\mubar{\bar\mu}
\def\dalpha{{\dot\alpha}}

\def\N{{\cal N}}
\def\M{{\cal M}}

\def\dbeta{{\dot\beta}}
\def\abar{\bar a}
\def\wbar{\bar w}
\def\trtwo{\tr^{}_2\,}

\def\bigL{{\bf L}}

\begin{document}

\addtolength{\baselineskip}{2pt}
\thispagestyle{empty}

\begin{flushright}
{\tt hep-th/9905209}\\
May 1999
\end{flushright}

\vspace{1cm}

\begin{center}
{\scshape\Large Summing the Instanton Series\\
\vspace{0.15cm}
in $\N=2$ Superconformal Large-$N$  QCD\\}

\vspace{1cm}

{\scshape Timothy J.~Hollowood$^{1,3}$,
Valentin V.~Khoze$^2$ and Michael
P.~Mattis$^1$}

\vspace{0.3cm}
$^1${\sl Theoretical Division T-8, Los Alamos National Laboratory,\\
Los Alamos, NM 87545, USA}\\
 {\tt pyth@schwinger.lanl.gov}, {\tt mattis@lanl.gov}\\

\vspace{0.2cm}
$^2${\sl Department of Physics, University of Durham,\\
Durham, DH1 3LE, UK}\\ {\tt valya.khoze@durham.ac.uk}\\

\vspace{0.2cm}
$^3${\sl Department of Physics, University of Wales Swansea,\\
Swansea, SA2 8PP, UK}\\

\vspace{1cm}

{\Large ABSTRACT}
\end{center}
\vspace{0.1cm}

\noindent 
We consider the multi-instanton collective coordinate integration
measure in $\N=2$ supersymmetric $SU(N)$ gauge theory with $N_F$
fundamental hypermultiplets. In the large-$N$ limit, at the
superconformal point where $N_F=2N$ and all VEVs are turned off, the
$k$-instanton moduli space collapses to a single copy of $AdS_5\times
S^1$. The resulting $k$-instanton effective measure is proportional to
$\sqrt{N}g^4\hat{\cal Z}_k^{(6)},$ where $\hat{\cal Z}_k^{(6)}$ is the
partition function of $\N=(1,0)$ SYM theory in six dimensions reduced
to zero dimensions. The multi-instanton can in fact be summed in
closed form.
As a hint of an AdS/CFT duality, with the usual relation between the gauge
theory and string theory parameters, this precisely matches  the
normalization of the charge-$k$ \hbox{D-instanton} measure in type IIB
string theory compactified to six dimensions on $K3$ with a vanishing
two-cycle. 

\newpage

Recently, it has been shown that the multi-instanton measure of $\N=4$
supersymmetric $SU(N)$ Yang-Mills theory, in the conformal case of vanishing VEVs, at
large $N$, has a precise correspondence with superstring theory on
$AdS_5\times S^5$ in the following sense \cite{lett,MO-III}:

(i) The effective large-$N$ collective coordinate space (`moduli
space') of the charge $k$ instanton
sector is $AdS_5\times S^5$.

(ii) The large-$N$ collective coordinate integration measure for $k$ instantons matches precisely the
measure for D-instantons in the string theory; in particular, it is
proportional to $\sqrt{N}g^8\hat{\cal Z}_k^{(10)},$ where
$\hat{\cal Z}_k^{(10)}$ is
the partition function for
$\N=(1,0)$ supersymmetric Yang-Mills (SYM) theory in ten dimensions reduced to zero dimensions.

(iii) Various correlation functions which have instanton
contributions match those calculated from string corrections to the
type IIB supergravity effective action as required by the AdS/CFT
correspondence \cite{Aharony:1999ti,MAL,GKP,WIT150,BG,BGKR}.

\noindent
These features add up to compelling circumstantial evidence in favor
of Maldacena's conjecture.

It is now tempting to apply the same large-$N$ instanton techniques to
theories without an obvious string theory dual in the hope that the
Yang-Mills instanton calculation will give some clues  about the
latter. From a Yang-Mills perspective a  simple variation on the
 $\N=4$ theory is   $\N=2$ supersymmetric $SU(N)$ gauge theory
 with $N_F$ fundamental hypermultiplets.
 As in the $\N=4$ case, we consider the theory with
vanishing VEVs. For $N_F=2N$ the theory then has non-trivial superconformal
invariance (just as the $\N=4$ theory); {\it a priori\/} we expect that
this case might be the most similar to its $\N=4$ cousin because the
coupling does not run and the instanton calculation can be justified
for small values of the 't Hooft coupling $g^2N$.

In analogy to the $\N=4$ model, we shall find that the multi-instanton
series may be summed in closed form, and the following features emerge:\footnote{Unlike 
the $\N=4$ case \cite{MO-III}, here we will not study correlation functions \it per
se \rm since a ``dictionary'' relating gauge-invariant operators on
the SYM and string sides of a proposed equivalence is missing in the
present case. To leading semiclassical order, $n$-point correlators
$\langle\Phi_1(x_1)\cdots\Phi_n(x_n)\rangle$ are straightforwardly obtained from the
collective coordinate measure obtained below, by inserting
$\Phi_1(x_1)\times\cdots\times\Phi_n(x_n)$ under the integral, where
each $\Phi_i$ is replaced by the corresponding classical expression
evaluated on the bosonic and/or fermionic collective coordinates of
the multi-instanton.}

(i) The effective large-$N$ moduli space of the charge $k$ instanton
sector is now $AdS_5\times S^1$.

(ii) The large-$N$ collective coordinate integration measure for $k$
instantons is proportional to $\sqrt{N}g^4\hat{\cal Z}_k^{(6)},$ where
$\hat{\cal Z}_k^{(6)}$ is the partition function for $\N=(1,0)$ SYM
theory in six dimensions reduced to zero dimensions. While an AdS/CFT
correspondence is not presently known for this $\N=2$ model, 
it is suggestive that this
expression precisely matches the normalization of the charge-$k$
\hbox{D-instanton} measure in type IIB string theory compactified to
six dimensions on $K3$ with a vanishing two-cycle \cite{GG}.  

In performing the $\N=2$ calculation,
we shall find that many of the required formulae and
notation can be adapted in an obvious way from the $\N=4$ theory,
discussed at great length in \cite{MO-III}. For this reason we shall
include only a paucity of details and refer frequently to this latter
reference. In the $\N=4$ theory the fermion and scalar fields come
with $SU(4)_R$ indices $A,B,\ldots=1,2,3,4$. For the present theory, in
order to describe the fermions and scalars of the vector multiplet we
only let $A,B,\ldots=1,2$. The scalar field is then the single complex field
$A^{12}=-A^{21}$ and the gluinos come in a pair $\lambda^A$, $A=1,2$,
which rotate into one another under $SU(2)_R.$ 
Virtually all the formulae of \cite{MO-III} for 
these fields in the background of an instanton
may be carried over. The new ingredients are the fundamental matter
fields. These, too, have been discussed at great length, in
\cite{KMS}, and we shall borrow profusely from this reference also. 

The charge $k$ instanton solution is described by the following
bosonic collective coordinates: the $k\times k$ dimensional Hermitian
matrices $a'_n$, $n=1,2,3,4$; the $N\times k$ dimensional matrices
$w_\aD$, where $\aD=1,2$ is a Weyl index; and their conjugates $\bar
w^\aD$. These are subject to quadratic ADHM constraints reviewed
below.  The gluino zero modes, $\lambda^A$, $A=1,2$, in the background
of the multi-instanton are described by the following Grassmann
collective coordinates: the $k\times k$ dimensional Hermitian matrices
$\M^{\prime A}_\alpha$, where $\alpha=1,2$; the $N\times k$
dimensional matrices $\mu^A$; and their conjugates $\bar\mu^A$. These
coordinates are subject to fermionic ADHM constraints which are
likewise reviewed below. Sometimes it is convenient to assemble
separately the bosonic and fermionic collective coordinates into
$(N+2k)\times k$ dimensional  matrices
$a=\begin{pmatrix} w_\aD\\ a'_{\alpha\aD}\end{pmatrix}$ and
$\M^A=\begin{pmatrix}\mu^A\\ \M^{\prime A}_\alpha\end{pmatrix}$, where
$a'_{\alpha\aD}=a'_n\sigma^n_{\alpha \aD}.$

In addition to the collective coordinates coming from the vector
multiplet, there are additional  fermionic (but not bosonic) collective
coordinates for the fundamental matter fields denoted $\K$ and $
\tilde\K$, which are $k\times N_F$ and $N_F\times k$ dimensional
matrices, respectively, of Grassmann numbers \cite{KMS}. These collective
coordinates account for the $2kN_F$ fundamental fermion zero modes in
the $k$ instanton background. Unlike the collective coordinates 
describing the vector multiplet there are no ADHM constraints 
on $\K$ and $\tilde\K$.

Of vital importance in constructing the instanton measure is the action
of the theory evaluated on the instanton solution, with all the
fermion modes `turned on'. 
In the case when the VEVs vanish, the action of the theory evaluated on
the multi-instanton solution has the form \cite{KMS}
\begin{equation}
\Skinst\ =\ {8\pi^2k\over g^2}\ -ik\theta\ +\ S^k_{\rm quad}\ ,
\elabel{Skinstdef}\end{equation}
where the term $\Skquad$ is a particular fermion quadrilinear term of
the form
\begin{equation}\Skquad\ =\ -{8\pi^2\over g^2}\,{\rm
tr}_k\,\Lambda_{\rm hyp}
\bigL^{-1}\Lambda_{\rm vec}\ .
\elabel{Skquadef}\end{equation}
Here we have defined the $k\times k$ matrix fermion bilinears
\begin{subequations}\begin{align}
&\Lambda_{\rm vec}={1\over2\sqrt2}(\bar\M^1\M^2-\bar\M^2\M^1)\ ,\\
&\Lambda_{\rm hyp}={\sqrt2\over8}\K\tilde\K\ 
\end{align}\end{subequations}
and $\BL$ is linear operator on $k\times k$ matrices defined in
\cite{KMS,MO-III}.
The existence of a four-fermion interaction \eqref{Skquadef}
between the fermionic collective coordinates played a vital r\^ole in
the $\N=4$ theory and we expect it to do so here as well.

The four-fermion interaction actually lifts all the fermion zero modes
except the 8 supersymmetric and superconformal adjoint zero modes $\zeta_\alpha^A$
and $\bar\eta^{\aD A}$ which will be explicitly defined later
(or see \cite{MO-III}). From the form of the quadrilinear action
\eqref{Skquadef}, we see that adjoint zero modes,
numbering $4kN$, and fundamental zero modes, numbering $2kN_F$, are
lifted in equal number. 
In any given $n$-point function, the $n$ insertions must, of course, 
saturate the integrals over the 
8 exact supersymmetric and superconformal adjoint zero modes but furthermore let us
suppose that they saturate the integrals over $n_a-8$
extra adjoint fermion zero modes and $n_f$ fundamental fermion
zero modes. We then have the following selection rule:
\begin{equation}
n_a-n_f=2k(2N-N_F)\ .
\elabel{select}
\end{equation}
{}From this selection rule it immediately follows that:

(i) Unless one requires $N_F\sim 2N$ in the large-$N$ limit, the only
non-vanishing correlators in this limit are ones where the number of
insertions grows with $N$. In this case the insertions will
necessarily modify the large-$N$ saddle-point equations derived below
(see for example the analysis of Ref.~\cite{gluinoc}).

(ii) It is obvious from Eq.~\eqref{select} that it is only in the conformal case 
$N_F=2N$ that a given correlation function gets a
contribution from each instanton number $k$ rather than from only one
value of $k$. Moreover, in the absence of VEVs, only in the conformal case 
can one choose an arbitrarily small coupling constant and hence rigorously
justify the instanton approximation (again, see Ref.~\cite{gluinoc}
for an analysis of what goes wrong with instanton physics when the
VEVs are zero but the $\beta$-function is not). 

\noindent
{}For these reasons, we concentrate henceforth on the conformal case $N_F=2N$.

The next ingredient needed in the analysis is
the integration measure $d\mu_{\rm vec}^k$ on the space of collective coordinates of the vector
multiplet. This has been determined in Refs.~\cite{meas1,meas2,KMS,MO-III}. It
is simply the `flat Cartesian measure' for the bosonic and fermionic
collective coordinates with explicit delta functions that impose the aforementioned
bosonic and fermionic ADHM constraints:\footnote{The
integrals over $k\times k$ matrix quantities $a'_n$ and $\M^{\prime
A}_\alpha$ and the arguments of the delta functions are defined with
respect to a basis of hermitian matrices $T^r$, $r=1,\ldots,k^2$,
normalized so that ${\rm tr}_k\,T^rT^s=\delta^{rs}$.}
\begin{equation}\begin{split}
&\int d\mu^k_{\rm vec}\ =\ {2^{-k^2/2}\over{\rm Vol}\,U(k)}
\int d^{4k^2} a' \
d^{2kN} \wbar \ d^{2kN} w  
\prod_{A=1,2}
\ d^{2k^2} \M^{\prime A}  
\ d^{kN} \mubar^A \ d^{kN} \mu^A\\
&\times \prod_{r=1,\ldots,k^2}\Big[
\prod_{c=1,2,3}
\delta\big(\tfrac12{\rm
tr}_k\,T^r(\trtwo\, \tau^c \abar a)\big)
\prod_{A=1,2}
\prod_{\aD=1,2}\delta\left({\rm tr}_k\,T^r(\Mbar^A a_\aD + \abar_\aD \M^A
)\right)\Big](\det_{k^2}\bigL)^{-1}\ .
\elabel{dmudef}\end{split}\end{equation}
The full measure is given by
\begin{equation}
\int d\mu^k_{\rm phys}=(C'_1)^k\int d\mu^k_{\rm vec}\times d\mu^k_{\rm hyp}\
,
\elabel{fullmeas}\end{equation}  
where $d\mu^k_{\rm hyp}$ is the measure for the fermionic collective
coordinates from the matter fields; this is simply
\begin{equation}
\int d\mu_{\rm hyp}^k=\pi^{-2kN_F}\int d^{kN_F}\K\,d^{kN_F}\tilde\K\ ,
\end{equation}
where the integrals $d^{kN_F}\K$ and $d^{kN_F}\tilde\K$ are defined as
separate integrals over each
component of the matrices $\K$ and $\tilde\K$ \cite{KMS}.
In \eqref{fullmeas} the constant $C'_1$ is determined by a detailed
comparison of this measure with the one instanton measure of Bernard
for $SU(N)$ \cite{Bernard} suitably generalized to an $\N=2$ theory with
fundamental hypermultiplets. This comparison gives
\begin{equation}
C'_1=2^{1/2}\pi^{-2N}\Lambda^{2N-N_F}\ ,
\end{equation}
where $\Lambda$ is the dynamical mass scale in the Pauli-Villars regularization scheme.

{}Following the same strategy as in \cite{KMS,MO-III}, we can bilinearize
the four fermion interaction by introducing a set of auxiliary
variables $\chi_a$, $a=1,2$, a two-vector of hermitian $k\times k$ matrices. In
addition, it is convenient to define the matrix of complex variables
\begin{equation}
\chi=\chi_1+i\chi_2\ .
\end{equation}
The transformation we need was established in \cite{KMS} and later
adapted to the $\N=4$ theory in \cite{MO-III}:
\begin{equation}\begin{split}
&(\det_{k^2}\BL)^{-1}\exp\,-S_{\rm quad}^k\\
 &=\ \pi^{-k^2}\int
d^{2k^2}\chi\exp\big[-{\rm tr}_k\,\chi_a\BL
\chi_a+\sqrt8\pi g^{-1}{\rm
tr}_k\,\Lambda_{\rm hyp}\chi+\sqrt8\pi g^{-1}{\rm
tr}_k\,\chi^\dagger\Lambda_{\rm vec}\big]. 
\elabel{E53}\end{split}\end{equation}
Conveniently,  the factor of $(\det_{k^2}\BL)^{-1}$ needed in this Gaussian
 rewriting is supplied by the flat measure, \eqref{dmudef}.
Next we transform to the so-called gauge invariant measure
\cite{MO-III}; this change of variables has the enormous calculational
 advantage that it allows us to integrate out explicitly both the
bosonic and fermionic ADHM constraints. The necessary manipulations
are detailed in \cite{MO-III}. The resulting measure involves the
bilinear bosonic variables
\begin{equation}
W_{\ \dbeta}^\dalpha=\bar w^\dalpha
 \,w_{\dbeta}\ ,\quad
W^0={\rm tr}_2\,W,\quad W^c={\rm
tr}_2\,\tau^cW, \ \ c=1,2,3\ ,
\elabel{Wdef}\end{equation}
where $W^0$ and $W^c$ are $k\times k$ hermitian matrices,
and a decomposition of the fermionic collective coordinates $\{\mu^A,\bar\mu^A\}$:
\begin{equation}
\mu^A=w_{\aD}\zeta^{\dalpha 
 A}+\nu^A,\qquad
\bar\mu^A=\bar\zeta^{
A}_\dalpha\bar w^\dalpha+\bar\nu^A\ ,
\elabel{zetadef}\end{equation}
where $\{\nu^A,\bar\nu^A\}$ lie in the orthogonal subspace to $w$ in
 the sense that
\begin{equation}
\bar w^\aD\nu^A=0\ ,\qquad \bar\nu^A w_\aD=0\ .
\elabel{nudef}\end{equation}
Here $\zeta^{\dalpha A}$ and $\bar\zeta^{A}_\dalpha$ are $k\times k$
matrices of Grassmann spinors.
The bosonic coordinates $W^c$, $c=1,2,3$ and the fermionic coordinates
$\bar\zeta^A_\aD$ are then explicitly integrated-out using, respectively, the
delta-functions in \eqref{dmudef} whose arguments are the bosonic and fermionic ADHM constraints.

As in the $\N=4$ model discussed in Ref.~\cite{MO-III}, the next stage
involves explicitly integrating out the $\nu^A$'s and
$\bar\nu^A$'s. Concentrating on the final term in the exponent in
Eq.~\eqref{E53}, one expands:
\begin{equation}
\Lambda_{\rm vec}\ =\ {1\over\sqrt{8}}(\bar\nu^1\nu^2-\bar\nu^2\nu^1)\
+\ \tilde\Lambda_{\rm vec}\ .
\elabel{eee}
\end{equation}
Here we have defined a $k\times k$ matrix fermion bilinear in the
remaining---unintegrated---fermionic collective coordinates:
\begin{equation}
\tilde\Lambda_{\rm vec}={1\over2\sqrt2}\big(\ 
\bar\zeta^1_\aD W^\aD_{\ \bD}\zeta^{\bD 2}-\bar\zeta^2_\aD W^\aD_{\
\bD}\zeta^{\bD 1}
+\{{\cal M}^{\prime\alpha 1},{\cal M}^{\prime 2}_\alpha\} \big)\ ,
\elabel{E52}\end{equation}
where $\bar\zeta^A_\aD$, in turn, is defined in terms of $\zeta^{\aD A}$ and
$\M_\alpha^A$ using the fermionic ADHM constraints.
Performing the integration over the $\nu^A$'s and $\bar\nu^A$'s then gives:
\begin{equation}
\int\prod_{A=1,2}d^{k(N-2k)}\nu^A\, d^{k(N-2k)}\bar\nu^A\,
\exp\big[\pi g^{-1}{\rm tr}_k
\,\chi^\dagger(\bar\nu^1\nu^2-\bar\nu^2\nu^1)\big]=\Big({\pi\over
g}\Big)^{2k(N-2k)}\, 
\big({\rm det}_{k}\chi^\dagger\big)^{2(N-2k)}\ .
\elabel{intoutnu}\end{equation}
This term will contribute to the saddle-point at large $N$. 

Next we turn to the fermionic collective coordinates in 
the matter sector. 
{}Focusing on the middle term in the exponent in Eq.~\eqref{E53} and
integrating out the $\K$ and $\tilde\K$ modes gives (for general $N_F$):
\begin{equation}
\int d^{kN_F}\K\,d^{kN_F}\tilde\K\,\exp\big[\tfrac12\pi g^{-1}
{\rm tr}_k\,\K\tilde\K\chi\big]=\Big({\pi\over
2g}\Big)^{kN_F} \big({\rm det}_{k}\chi\big)^{N_F}\ .
\elabel{intoutk}\end{equation}
Putting together all the previous steps, 
we now have a form for the measure which is amenable to a large-$N$ limit:
\begin{equation}\begin{split}
&\int\dmuphys\,e^{-\Skinst}\ =\ {2^{5k^2/2-kN_F}\pi^{k(2N-N_F)-4k^2}(C'_1)^kc_{k,N}\over
g^{k(2N-N_F)-4k^2}\,{\rm Vol}\,\U(k)}e^{2\pi ik\tau}\\
&\qquad\qquad\qquad\times\,\int d^{k^2}W^0\,d^{4k^2}a'\,
d^{2k^2}\chi\,\prod_{A=1,2}d^{2k^2}\M^{\prime A}\,
d^{2k^2}\zeta^A\\
&\times\
\big(\det_{2k} W\big)^{N-2k}
\big({\rm det}_{k}\chi^\dagger\big)^{2(N-2k)}
\big({\rm det}_{k}\chi\big)^{N_F}
\,\exp\big[\sqrt8\pi g^{-1}\,{\rm tr}_k\,\chi^\dagger\tilde
\Lambda_{\rm vec}
\ -\ {\rm tr}_k\,\chi_a\BL\chi_a\big]\, .
\elabel{gaugeinvmeas}
\end{split}\end{equation}
Here the constant $c_{k,N}$ comes from switching to gauge-invariant
variables, and is given by \cite{MO-III}:
\begin{equation}
c_{k,N}\ =\
{2^{2kN-4k^2+k}\pi^{2kN-2k^2+k}\over\prod_{i=1}^{2k}(N-i)!}\ .
\elabel{E37.1}\end{equation} 

As in \cite{MO-III}, in order to facilitate taking the large-$N$ limit,
it is useful to re-scale $\chi$:
\begin{equation}
\chi\rightarrow\sqrt N\chi\ .
\end{equation}
With this rescaling, the last term in the exponent of
Eq.~\eqref{gaugeinvmeas} now scales like $N$, hence contributes to the
large-$N$ saddle-point equations (as do the three determinantal
factors preceding the exponential).
Collecting all the bosonic terms which have the form of an exponent of
``$N$ times something'' and specializing once again to $N_F=2N$ defines the ``large-$N$ effective
instanton action'':
\begin{equation}S_{\rm
eff}=-\log{\rm det}_{2k}W-2\log{\rm det}_{k}\chi^\dagger 
-2\log{\rm det}_{k}\chi+{\rm
tr}_k\,\chi_a\BL\chi_a\, . 
\elabel{E57}\end{equation}

We now turn to the solution of the large-$N$ saddle-point
equations. These are the coupled Euler-Lagrange equations that come
from extremizing $S_{\rm eff}$ with respect to the remaining bosonic
collective coordinates in the problem.
The analysis of these coupled equations, while rather involved, is
virtually identical to the $\N=4$ model of Ref.~\cite{MO-III}; hence
we suppress the calculational details. As in that model, the
saddle-point solution
that is dominant in the large-$N$ limit is given by:
\begin{equation}
W^0=2\rho^2\,1_{\sst[k]\times[k]}\ ,\qquad
\chi_a=\rho^{-1}\sfc_a\,1_{\sst[k]\times[k]}\ ,\qquad
a'_n=-\com_n\,1_{\sst[k]\times[k]}\ ,
\elabel{specsol}
\end{equation}
where $\sfc_a$ is a unit 2-vector. It is convenient to parametrize
$\sfc_a$ by the phase angle $\phi$:
\begin{equation}
\sfc_1+i\sfc_2=e^{i\phi}\ .
\end{equation}
The interpretation of the solution
is as in \cite{MO-III} with the $S^5$ replaced by $S^1$, parametrized
by $\sfc_a$. In other words, the solution is parametrized by a point
$(X_n,\rho,\phi)$ on
$AdS_5\times S^1$. As in the $\N=4$ case, the solution \eqref{specsol}
is not actually the most general solution to the saddle-point 
equations; there are additional flat directions (moduli). However, we shall find in a completely
analogous way that
the integrals over the additional moduli are actually convergent;
as such, they are more conveniently viewed as fluctuations around, rather
than facets of, the maximally symmetric saddle-point solution \eqref{specsol}.
 
The next stage of the analysis is to expand the effective action in
the fluctuations out to sufficient order to ensure the convergence of
the integrals over the fluctuations. Fortunately, the analysis need
not be repeated because our effective action \eqref{E57} is, up to a
constant, simply the $\N=4$ effective action of \cite{MO-III} with the
replacements $\chi_{12}\to\chi/\sqrt8$,
$\chi_{34}\to\chi^\dagger/\sqrt8$ and all other components of
$\chi_{AB}$ set to zero. The fluctuations in $W^0$ are integrated out
at Gaussian order to leave integrals over the remaining fluctuations
in $\hat a'_n$ and $\hat\chi_a$, the traceless parts, that are lifted
at quartic order; the terms beyond quartic order are then formally
suppressed by (fractional) powers of $1/N$, and may be dropped. The
`action' for these quartic fluctuations is precisely the action for
six-dimensional Yang-Mills theory with gauge field
\begin{equation}
A_\mu=N^{1/4}\big(\rho^{-1}\hat a'_n,\rho\hat\chi_a\big)\ ,\quad
\mu=1,\ldots,6\ ,
\elabel{amudef}\end{equation}
dimensionally reduced to $0$ dimensions, i.e. with all derivatives
set to zero:
\begin{equation}
S(A_\mu)=NS_{\rm eff}=-\tfrac12{\rm tr}_k\,\left[A_\mu,A_\nu\right]^2+\cdots \ .
\elabel{actymb}
\end{equation}
In \eqref{amudef} the hats on the variables imply that they are
traceless $k\times k$ matrices, i.e.~$SU(k)$-valued rather than $U(k)$. 

Not surprisingly the coupling to the fermions completes the
six-dimensional theory to an $\N=(1,0)$ supersymmetric gauge theory
in six dimensions dimensionally reduced to $0$ dimensions. It is easy to see, for instance, 
that this theory has the requisite number of supersymmetries. The eight
dimensional fermion field of the $\N=(1,0)$ theory has components
\begin{equation}
\Psi=\sqrt{\pi\over2g}\,N^{1/8}e^{-i\phi/2}\big(\rho^{-1/2}
\hat{\cal M}^{\prime A}_\alpha\,,\,
\rho^{1/2}\hat\zeta^{\aD A}\big)\ ,
\elabel{psidef}
\end{equation}
In terms of which the coupling to the vector completes the action
\eqref{actymb} to
\begin{equation}
S(A_\mu,\Psi)\ =\ 
-\tfrac12{\rm tr}_k\,\left[A_\mu,A_\nu\right]^2 
+{\rm
tr}_k\,\bar\Psi\Gamma_\mu\left[A_\mu,\Psi\right]\ .
\elabel{sukpart}
\end{equation}
In \eqref{psidef} we have
used the decomposition of the fermionic collective coordinates
\begin{subequations}
\begin{align}
\M^{\prime A}_\alpha&=4\xi^A_\alpha1_{\sst
[k]\times[k]}+4\hat a'_{\alpha\aD}\bar\eta^{\aD A}+\hat\M^{\prime A}_\alpha\ ,\elabel{sosm}\\
\zeta^{\aD A}&=4\bar\eta^{\aD A}1_{\sst [k]\times[k]}+\hat\zeta^{\aD
A}\ ,
\end{align}
\end{subequations}
where $\xi^A_\alpha$ and $\bar\eta^{\aD A}$ are the 
4 supersymmetric and 4 superconformal fermion
modes, respectively. In \eqref{sukpart} the $\Gamma_\mu$ are a
representation of the six-dimensional Clifford algebra.

{}Finally our effective measure for the $k$ instantons has the form
\begin{equation}\begin{split}
\int\dmuphys\,e^{-\Skinst}\ \underset{N\rightarrow\infty}=&
\ {N^{1/2}g^{4}\over
2^{9k^2/2-k/2+12}\,\pi^{5k^2/2+4}\,{\rm Vol}\,U(k)}k^{-1}e^{2\pi ik\tau}\\
&\times\ \int\,d\rho\,d^4\com\, d\phi\,\rho^{-5}\,e^{4i\phi}\,\prod_{A=1,2}d^2\xi^A 
d^2\bar\eta^A
\cdot\hat{\cal Z}^{(6)}_k\ ,
\elabel{hello}\end{split}\end{equation}
where $\hat{\cal Z}_k^{(6)}$ is the partition function of an ${\cal N}=(1,0)$
supersymmetric $\SU(k)$ gauge theory in six dimensions dimensionally reduced
to zero dimensions:
\begin{equation}
\hat{\cal Z}_k^{(6)}\ =\ \int_{SU(k)}\, 
d^{6}A\, d^{8}\Psi\,e^{-S(A_\mu,\Psi)}\ .
\elabel{part}
\end{equation}
When integrating expressions which are independent of the $\SU(k)$
degrees-of-freedom, $\hat{\cal Z}_k^{(6)}$ is simply an overall constant
factor that was evaluated in \cite{KNS}. In our notation\footnote{We
have written the result in a way which allows an easy comparison with
\cite{KNS}. The factors of $\sqrt{2\pi}$ and $\sqrt2$ arise, respectively, from the
difference in the definition of the bosonic integrals and the normalization
of the generators: we have ${\rm tr}_k\,T^rT^s=\delta^{rs}$ rather
than $\tfrac12\delta^{rs}$. The
remaining factors are the result of \cite{KNS}. We have also used ${\rm Vol}\,U(k)
={2^k\pi^{k(k+1)/2}/\prod_{i=1}^{k-1}i!}$.}
\begin{equation}
\hat{\cal
Z}^{(6)}_k=\big(\sqrt{2\pi}\big)^{6(k^2-1)}\big(\sqrt2\big)^{(8-6)(k^2-1)}\cdot
{2^{k(k+1)/2}\pi^{(k-1)/2}\over2\sqrt
k\prod_{i=1}^{k-1}i!}\cdot{1\over k^2}\ .
\elabel{parte}\end{equation}

In summary, the effective large-$N$ collective coordinate measure has the following
simple form:
\begin{equation}
\int\dmuphys\,e^{-\Skinst}\ \underset{N\rightarrow\infty}=\ {\sqrt Ng^4\over
2^{17}\pi^{15/2}}\,k^{-7/2}e^{2\pi ik
\tau}
\,\int\,
{d^4\com\,d\rho\over\rho^5}\, d\phi\,e^{4i\phi}\,\prod_{A=1,2}d^2\xi^A 
d^2\bar\eta^A\ .
\elabel{endexpcc}
\end{equation}
This has a remarkable similarity to the form of
the $\N=4$ measure (see Eq.~(5.45) in \cite{MO-III}). Apart from the
differences in the overall numerical factors, the integral
over $S^5$ is replaced by $S^1$ and the $\N=4$ measure involves, in addition, the
factor $\sum_{d|k}d^{-2}$, the sum over the integer divisors of
$k$. Notice that the $\sqrt N$ dependence and factor of $k^{-7/2}$ is the same in both cases.
We can sum the instanton series in this case in terms of a generalized
zeta-function:\footnote{We thank Michael Nieto for pointing this out \cite{GR}.}
\begin{equation}
\sum_{k=1}^\infty k^{-7/2}e^{2\pi ik\tau}=e^{2\pi i\tau}\Phi(e^{2\pi
i\tau},\tfrac72,1)\ .
\end{equation}

The appearance of the phase in 
\eqref{endexpcc} implies a selection
rule in order that correlation functions are non-vanishing. In fact this is
exactly the selection rule \eqref{select} since, as is clear from the
couplings of the fermion bilinears to $\chi$ in \eqref{E53}, each insertion of an
adjoint fermion mode, other than the 8 exact modes, (numbering
$n_a-8$) implies an
insertion of $e^{i\phi/2}$ and each insertion of a fundamental fermion
mode (numbering $n_f$) implies an insertion of $e^{-i\phi/2}$. Using this selection rule
we see that the simplest 
kind of non-vanishing correlation functions would involve
operators that saturate the 8 supersymmetric and superconformal zero
modes and also 8 fundamental fermion modes, i.e.~$n_a=8$ and
$n_f=8$. 

In the $\N=4$ theory the large-$N$ instanton measure could be
interpreted directly as the D-instanton measure for type IIB string
theory on $AdS_5\times S^5$. The natural question to ask is whether
there is any similar interpretation for the measure \eqref{endexpcc}
in terms of D-instantons in six dimensions. Remarkably there is such a
relation arising from the 
type IIB superstring compactified on $K3$ \cite{GG}. 
D-instantons arise from wrapping the world-volume of the
D-string in ten dimensions on an $S^2$ in $K3$ at a special point in
the latter where the $S^2$ vanishes. More generally, 
Green and Gutperle \cite{GG} were able to deduce
the normalization of the measure for D-instantons in various
dimensions (4, 6 and 10) from
the exact expressions for certain
higher-derivative terms in the supergravity effective
actions in those dimensions. For arbitrary
dimension $d$, the leading order piece in the weak
coupling expansion has the generic form:
\begin{equation}
(\alpha')^{-1}(S_k)^{a_d}e^{-2\pi(S_k-ik\chi)}\hat{\cal Z}_k^{(d)}\big(1+{\cal
O}(S_k^{-1})\big)\ ,
\end{equation}
where $S_k=ke^{-\phi}$ is the $k$-instanton action, where $\phi$ is
the expectation value of the 
dilaton, the $NS$-$NS$ scalar, and $\chi$ is the expectation value of
the $R$-$R$ scalar. In the above $a_d$
is a constant that depends upon the dimension and $\hat{\cal Z}_k^{(d)}$ is the
partition function of $SU(k)$ minimally supersymmetric gauge theory in
$d$-dimensions dimensionally reduced to zero dimensions, 
as in \eqref{part}, but with a different normalization:
\begin{equation}
\hat{\cal Z}_k^{(10)}=\sum_{d|k}d^{-2}\ ,\qquad 
\hat{\cal Z}_k^{(6)}=\hat{\cal Z}_k^{(4)}=k^{-2}\ .
\end{equation}
Hence in six dimensions, where $a_6=-\tfrac32$, the leader order 
behaviour is 
\begin{equation}
(\alpha')^{-1}e^{3\phi/2}k^{-7/2}\exp(2\pi ik(ie^{-\phi}+\chi))\ .
\elabel{mea}\end{equation}
With the usual identification of coupling constants of the
Yang-Mills theory at large $N$ and the string theory parameters;
namely $g=\sqrt{4\pi e^{\phi}}$, $\theta=2\pi\chi$ and $(\alpha')^{-1}\sim \sqrt{g^2N}$,
\eqref{mea} goes like
\begin{equation}
\sqrt Ng^4 k^{-7/2}e^{2\pi ik\tau}\ ,
\end{equation}
which matches the normalization of \eqref{endexpcc} exactly.

It remains to be seen whether the relation of the D-instanton measure
of the type IIB string compactified on the special $K3$ and the large
$N$ multi-instanton measure in $\N=2$
$SU(N)$ gauge theory with $2N$ fundamental
hypermultiplets is a reflection of an AdS/CFT-type duality. However, we
should make the following cautionary remark. It is believed that for a conformal
gauge theory to admit a SUGRA dual in the large $N$ limit the
two coefficients of the four-dimensional anomaly, denoted as $a$ and
$c$ in \cite{Anselmi:1997am}, have to be equal at leading order in
$1/N$ \cite{Henningson:1998gx,Gubser:1998vd}. 
For the $\N=2$ conformal theory that we are considering
this is not the case since 
$a=\tfrac1{24}(7N^2-5)$ and $c=\tfrac16(2N^2-1)$
\cite{Anselmi:1998te}.\footnote{This was pointed out to us by Michael
Gutperle \cite{Gutperle:1999xu}, who recently considered D-instantons and an
AdS/CFT correspondence in an $\N=2$ conformal theory  
involving the gauge group $Sp(N)$; however, in this case $a=c$ and a SUGRA dual exists.} 
This implies that the dual, if it exists, is a
string theory which does not have a SUGRA approximation. 

We would like to thank Matt Strassler for encouraging us to
investigate theories other than $\N=4$ Yang-Mills using large-$N$
instanton techniques. We would also like to thank Michael Green and
Michael Gutperle and Michael Nieto for very useful comments and suggestions.
VVK and MPM acknowledge a NATO Collaborative Research Grant,
TJH and VVK acknowledge the TMR network grant FMRX-CT96-0012.

\end{document}